\documentclass{article}
\usepackage{spconf,amsmath,graphicx}
\usepackage{tabto}
\usepackage{amsfonts}
\usepackage{lipsum}

\title{LaSAFT: Latent Source Attentive Frequency Transformation for Conditioned Source Separation}

\name{Woosung Choi$^{\star}$ \thanks{This work was supported by the National Research Foundation of Korea(NRF) grant funded by the Korea government(MSIT)(No. 2020R1A2C1012624, 2019R1A6A3A13095526).}, Minseok Kim$^{\star}$, Jaehwa Chung$^{\dagger}$, Soonyoung Jung$^{\star}$}

\address{$^{\star}$ Department of Computer Science, Korea University \\
			    $^{\dagger}$ Department of Computer Science, Korea National Open University}

\begin{document}
%
\maketitle
\begin{abstract}

Recent deep-learning approaches have shown that Frequency Transformation (FT) blocks can significantly improve spectrogram-based single-source separation models by capturing frequency patterns. The goal of this paper is to extend the FT block to fit the multi-source task. We propose the Latent Source Attentive Frequency Transformation (LaSAFT) block to capture source-dependent frequency patterns. We also propose the Gated Point-wise Convolutional Modulation (GPoCM), an extension of Feature-wise Linear Modulation (FiLM), to modulate internal features. By employing these two novel methods, we extend the Conditioned-U-Net (CUNet) for multi-source separation, and the experimental results indicate that our LaSAFT and GPoCM can improve the CUNet's performance, achieving state-of-the-art SDR performance on several MUSDB18 source separation tasks.

\end{abstract}
\section{Introduction}
\label{sec:intro}


Most of the deep learning-based models for Music Source Separation (MSS) are dedicated to a single instrument.
However, this approach forces us to train an individual model for each instrument. Besides, trained models cannot use the commonalities between different instruments.
A simple extension to multi-source separation is to generate several outputs at once. For example, models proposed in \cite{dilatedlstm, d3net} generate multiple outputs.
Although it shows promising results, this approach still has a scaling issue: the number of heads increases as the number of instrument increases, leading to (1) performance degradation caused by the shared bottleneck, (2) inefficient memory usage.

Adopting conditioning learning \cite{cunet, meta} is a useful alternative. It can separate different instruments with the aid of the control mechanism. Since it does not need a multi-head output layer, there is no shared bottleneck.
For example, the Conditioned-U-Net (CUNet) \cite{cunet} extends the U-Net \cite{medical_unet, svs_unet} by exploiting Feature-wise Linear Modulation (FiLM) \cite{film}.
It takes as input the spectrogram of a mixture and a control vector that indicates which instrument we want to separate and outputs the estimated spectrogram of the target instrument.

Meanwhile, recent spectrogram-based methods for Singing Voice Separation (SVS) \cite{tfctdf} or Speech Enhancement (SE) \cite{phasen} employed Frequency Transformation (FT) blocks to capture frequency patterns. 
Although stacking 2-D convolutions has shown remarkable results \cite{svs_unet,mmdenselstm}, it is hard to capture long-range dependencies along the frequency axis for fully convolutional networks with small sizes of kernels. FT blocks, which have fully-connected layers applied in a time-distributed manner, are useful to this end.
Both models designed their building block to have a series of 2-D convolution layers followed by an FT block, reporting the state-of-the-art performance on SVS and SE, respectively.

In this paper, we aim to exploit FT blocks in a CUNet architecture.
However, merely injecting an FT block to a CUNet does not inherit the spirit of FT block (although it does improve Source-to-Distortion Ratio (SDR) \cite{bss} performance by capturing common frequency patterns observed across all instruments). 
We propose the Latent Source-Attentive Frequency Transformation (LaSAFT), a novel frequency transformation block that can capture instrument-dependent frequency patterns by exploiting the scaled dot-product attention \cite{transformer}.
We also propose the Gated Point-wise Convolutional Modulation (GPoCM), a new modulation that extends the Feature-wise Linear Modulation (FiLM) \cite{film}.
Our CUNet with LaSAFT and GPoCMs outperforms the existing methods on several MUSDB18 \cite{musdb18} tasks. Our ablation study indicates that adding LaSAFT or replacing FiLMs with GPoCMs improves separation quality.

Our contributions are three-fold as follows:

\begin{itemize}
    \item We propose LaSAFT, an attention-based novel frequency transformation block that captures instrument-dependent frequency patterns.
    \item We propose GPoCM, an extension of FiLM, to modulate internal features for conditioned source separation.
    \item Our model achieves state-of-the-art performance on several MUSDB18 tasks. We provide an ablation study to investigate the role of each component.
\end{itemize}

\section{Baseline Architecture}
\label{sec:baseline}
The baseline is similar to the CUNet \cite{cunet}. It consists of (1) the Conditioned U-Net and (2) the Condition Generator.

\begin{enumerate}
    \item \textbf{The Conditioned U-Net} is a U-Net  \cite{medical_unet} which takes a mixture spectrogram as input and outputs the estimated target spectrogram. It applies FiLM layers to modulate intermediate features with condition parameters generated by the condition generator.
    \item \textbf{The Condition Generator} takes as input a condition vector and generates condition parameters. A condition vector is a one-hot encoding vector that specifies which instrument we want to separate.
\end{enumerate}

\subsection{The Conditioned U-Net}
\begin{figure}[htb]

  \centering
  {\includegraphics[width=8.5cm]{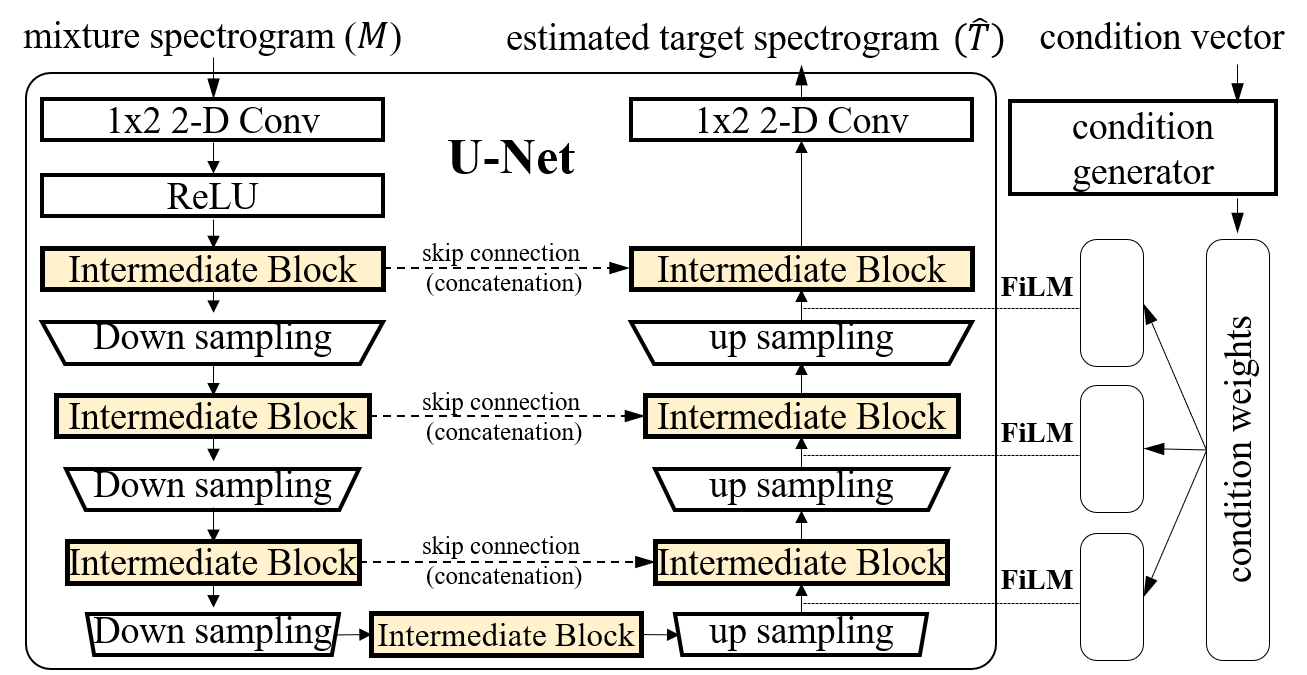}}

    \caption{The baseline architecture}
    \label{fig:baseline}
\end{figure}

We extend the U-Net architecture used in \cite{tfctdf}, on which a state-of-the-art singing voice separation model is based.
We first describe the common parts between ours and the original \cite{tfctdf}.
As shown in the left part of Fig. \ref{fig:baseline}, it consists of an encoder and a decoder: the encoder transforms the input mixture spectrogram $M$ into a downsized spectrogram-like representation. The decoder takes it and returns the estimated target spectrogram $\hat{T}$.
It should be noted that spectrogram $M$ and $\hat{T}$ are complex-valued adopting the Complex-as-Channel (CaC) separation method \cite{tfctdf}.
In CaC, we view real and imaginary as separate channels. Thus, if the original mixture waveform is $c$-channeled  (i.e., $c=2$ for stereo), then the number of channels of $M$ and $\hat{T}$ is ($2c$).

\noindent
There are four types of components in the structure:

\noindent
\textbf{\textit{1 X 2 Convolution Layers}} are used for adjusting the number of channels. To increase the number of channels, we apply a $1\times 2$ convolution with $\bar{C}$ output channels followed by ReLU \cite{relu} activation to the given input $M$. Intermediate layers keep the number of channels $\bar{C}$.
To restore the original number of channels, we apply another $1\times 2$ convolution with $(2c)$ output channels to the last intermediate block's output. Since the target spectrogram is complex-valued, we do not apply any activation functions.

\noindent
\textbf{\textit{An Intermediate Block}} transforms an input spectrogram-like tensor into an equally-sized tensor.
For each block in the baseline, we use a Time-Frequency Convolution \cite{tfctdf} (TFC), a block of densely connected 2-D convolution layers \cite{densenet}.
We denote the number of intermediate blocks in the encoder by $L$. The decoder also has $L$ blocks. There is an additional block between them. 

\noindent
\textbf{\textit{A Down/Up-sampling Layer}} halves/doubles the scale of an input tensor. We use a strided/transposed-convolution.

\noindent
\textbf{\textit{Skip Connections}} concatenate output feature maps of the same scale between the encoder and the decoder. They help the U-Net recover fine-grained details of the target.

Unlike in the original U-Net \cite{tfctdf}, we modulate internal features in the decoder by applying FiLM layers, as shown in the right part in Fig. \ref{fig:baseline}.
Applying a FiLM is an effective way to condition a network, which applies the following operation to intermediate feature maps.
We define a Film layer as follows:

\begin{equation}
    FiLM(X^{i}_{c}|\gamma_{c}^{i},\beta_{c}^{i}) =  \gamma_{c}^{i} \cdot X^{i}_{c} + \beta_{c}^{i} 
\end{equation}

\noindent
where $\gamma_{c}^{i}$ and $\beta_{c}^{i}$ are parameters generated by the condition generator, and $X^{i}$ is the output of the $i^{th}$ decoder's intermediate block, whose subscript refers to the $c^{th}$ channel of $X$. 





\subsection{The Condition Generator}

The condition generator is a network that predicts condition parameters $\gamma=(\gamma_{1}^{1}, ..., \gamma_{\bar{C}}^{L})$ and $\beta=(\beta_{1}^{1}, ..., \beta_{\bar{C}}^{L})$. 
Our condition generator is similar to `Fully-Connected Embedding' of the  CUNet \cite{cunet} except for the usage of the embedding layer. 
It takes as input the one-hot encoding vector $z \in \{0,1\}^{|\mathcal{I}|}$ that specifies which one we want to separate among $|\mathcal{I}|$ instruments. The condition generator projects $z$ into $e_{z} \in \mathbb{R}^E$, the embedding of the target instrument, where $E$ is the dimension of the embedding space. 

It then applies a series of fully connected (i.e., linear or dense) layers, which doubles the dimension. We use ReLU \cite{relu} as the activation function for each layer, and apply a dropout with $p=0.5$ followed by a Batch Normalization (BN) \cite{bn} to the output of each last two layers.
The last hidden units are fed to two fully-connected layers to predict $\gamma=(\gamma_{1}^{1}, ..., \gamma_{\bar{C}}^{L}) \in \mathbb{R}^{|\gamma|}$ and $\beta=(\beta_{1}^{1}, ..., \beta_{\bar{C}}^{L})\in \mathbb{R}^{|\beta|}$, where $|\gamma|=|\beta|=\bar{C}L$.




\section{Proposed Methods}

\subsection{Latent Source Attentive Frequency Transformation}

We introduce the Time-Distributed Fully-connected layers (TDF) \cite{tfctdf}, an existing FT, and then propose our LaSAFT.

\subsubsection{TDF: a Frequency Transformation Block}

A TDF is a series of two fully-connected layers.
Suppose that the 2-D convolutional layer (e.g., TFC \cite{tfctdf}) takes an input representation $X\in\mathbb{R}^{\bar{C}\times T \times F}$ and outputs equally-sized tensor $X'$.
A TDF is applied separately and identically to each frame (i.e., $X'[i,j,:]$) in a time-distributed fashion. 
Each layer is defined as consecutive operations: a fully-connected layer, BN, and ReLU. 
The number of hidden units is defined as $\lfloor F^{(l)}/bf \rfloor$, where we denote the bottleneck factor by $bf$. 

\subsubsection{Extending TDF to the Multi-Source Task}

Although injecting TDFs to the baseline also improves the SDR performance by capturing the common frequency patterns observed across all instruments (see \S \ref{sec:exp}), it does not inherit the spirit of TDF. To this end, we propose the Latent Source Attentive Frequency Transformation (LaSAFT) by adopting the scaled dot-product attention mechanism \cite{transformer}.

\begin{figure}[htb]

  \centering
  {\includegraphics[width=8.5cm]{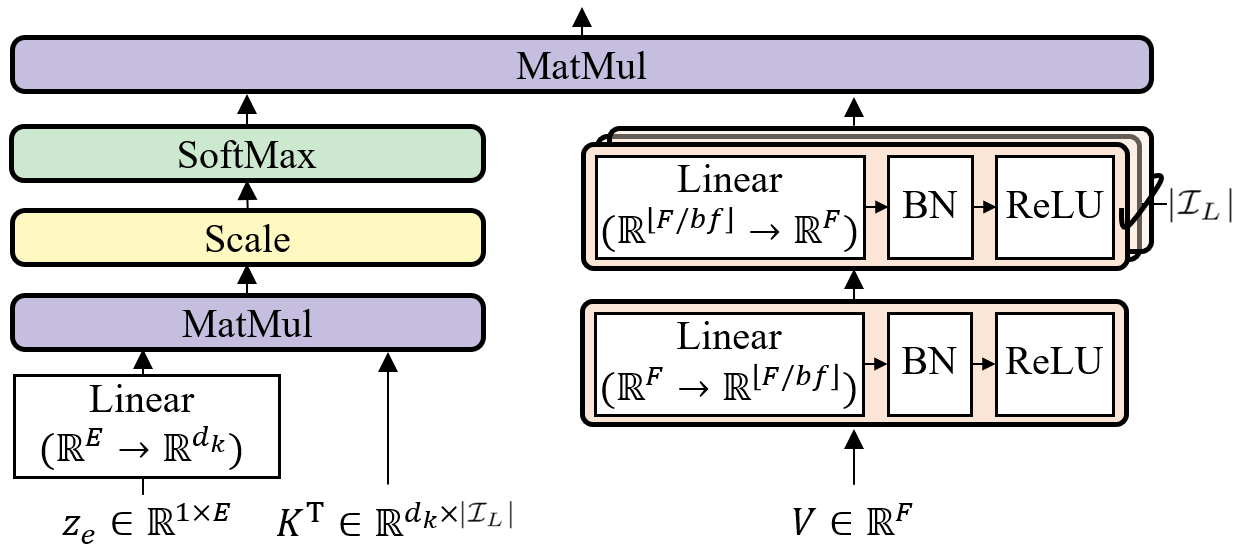}}

    \caption{Latent Source Attentive Frequency Transformation}
    \label{fig:lasaft}

\end{figure}

We first duplicate $|\mathcal{I}_L|$ copies of the second layer of the TDF, as shown in the right side of Fig. \ref{fig:lasaft}, where $|\mathcal{I}_L|$ refers to the number of \textit{latent instruments}. $|\mathcal{I}_L|$ is not necessarily the same as $\mathcal{I}$ for the sake of flexibility.
For the given frame $V\in \mathbb{R}^F$, we obtain the $|\mathcal{I}_L|$ latent instrument-dependent frequency-to-frequency correlations, denoted by $V'\in \mathbb{R}^{F \times |\mathcal{I}_L|}$.
We use components on the left side of Fig. \ref{fig:lasaft} to determine how much each \textit{latent source} should be attended. LaSAFT takes as input  the instrument embedding $z_e \in \mathbb{R}^{1 \times E}$. 
It has a learnable weight matrix $K\in \mathbb{R}^{ |\mathcal{I}_L| \times d_{k}}$, where we denote the dimension of each instrument's hidden representation by $d_{k}$.
By applying a linear layer of size $d_{k}$ to $z_e$, we obtain $Q \in \mathbb{R}^{d_{k}}$.
We now can compute the output of the LaSAFT as follows:

\begin{equation}
    Attention(Q,K,V') = softmax(\frac{QK^{T}}{\sqrt{d_{k}}})V'
\end{equation}

We apply a LaSAFT after each TFC in the encoder and after each Film/GPoCM layer in the decoder.
We employ a skip connection for the final output of each block (i.e., it  outputs $X'+lasaft(X')$, where $X'$ is an input of the $lasaft$).

\subsection{Gated Point-wise Convolutional Modulation}
\label{sec:format}

Before we describe the Gated Point-wise Convolutional Modulation (GPoCM), we first introduce the Point-wise Convolutional Modulation (PoCM).

\subsubsection{Point-wise Convolutional Modulation}

\begin{figure}[htb]

  \centering
  {\includegraphics[width=8.5cm]{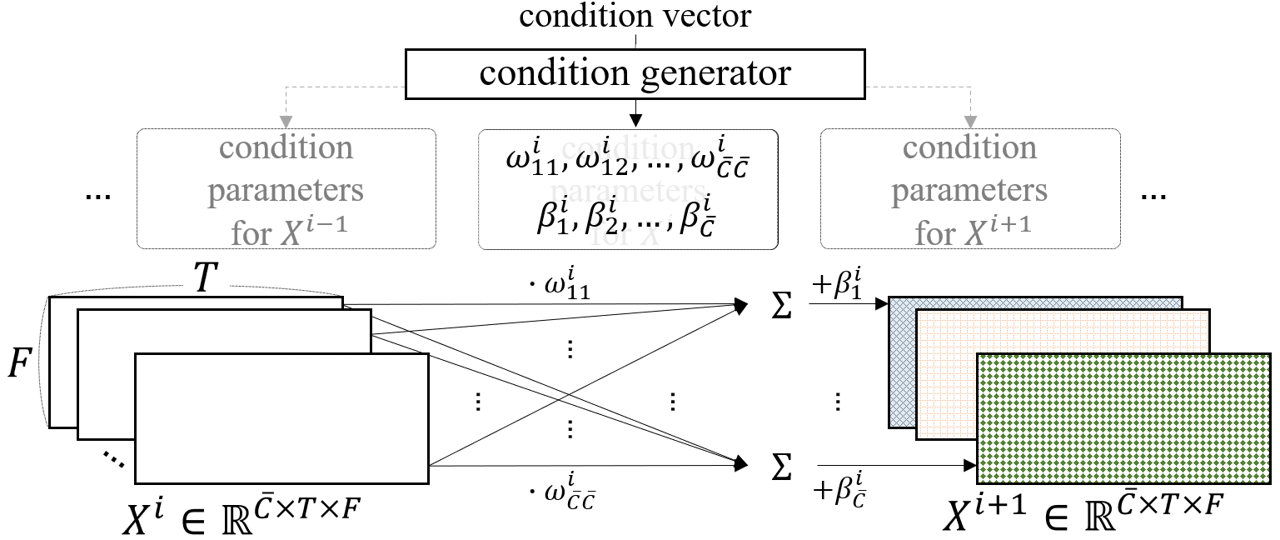}}

    \caption{PoCM layers}
    \label{fig:pocm}

\end{figure}

The PoCM is an extension of FiLM. While FiLM does not have inter-channel operations, PoCM has them as follows:

\begin{equation}
    PoCM(X^{i}_{c}|\omega_{c}^{i},\beta_{c}^{i}) = \beta_{c}^{i} +   \sum_{j}{\omega_{cj}^{i} \cdot X^{i}_{j}} 
\end{equation}

\noindent
where $\omega_{c}^{i}=(\omega_{c1}^{i}...,\omega_{c\bar{C}}^{i})$ and $\beta_{c}^{i}$ are condition parameters, and $X^{i}$ is the output of the $i^{th}$ decoder's intermediate block, as shown in Fig. \ref{fig:pocm}. Since this channel-wise linear combination can also be viewed as a point-wise convolution, we name it as PoCM. With inter-channel operations, PoCM can modulate features more flexibly and expressively than FiLM.

\subsubsection{Gated Point-wise Convolutional Modulation}

In the decoder, we use the following `Gated PoCM(GPoCM)' instead, as follows: 
\begin{equation}
    GPoCM(X^{i}_{c}|\omega_{c}^{i},\beta_{c}^{i}) = \sigma(PoCM(X^{i}_{c}|\omega_{c}^{i},\beta_{c}^{i})) \odot X^{i}_{c}
\end{equation}
\noindent
where $\sigma$ is a sigmoid and $\odot$ means the Hadamard product. 

\section{Experiment}
\label{sec:exp}

\subsection{Experiment Setup}

We use MUSDB18 dataset \cite{musdb18}, which contains 86 tracks for training, 14 tracks for validation, and 50 tracks for test. Each track is stereo, sampled at 44100 Hz, consisting of the mixture and four sources (i.e., $|\mathcal{I}|$=4): vocals, drums, bass, and other.

We train models using Adam \cite{adam} with learning rate  $lr \in \{0.0005, 0.001\}$ depending on model depth. 
Each model is trained to minimize the mean squared error between the ground-truth STFT output and the estimated.
For validation, we use mean absolute error of the target signal and the estimated. We apply data augmentation \cite{blend} on the fly to obtain mixture clips comprised of sources from different tracks.
We use SDR \cite{bss} for the evaluation metric, by using the official tool\footnote{https://github.com/sigsep/sigsep-mus-eval} for MUSDB. 
We use the median SDR value over all the test set tracks for each run and report the mean SDR over three runs.
More details are available online\footnote{https://github.com/ws-choi/Conditioned-Source-Separation-LaSAFT}.

\subsection{Results}

We perform an ablation study to validate the effectiveness of the proposed methods compared with the baseline.
In every model in the Table \ref{table:ablation}, we use an FFT window size of 2048 and hop size of 1024 for spectrogram estimation.
The configuration of the baseline (FiLM CUNET) is as follows: we use 7-blocked CUNet (i.e., $L=3$), and we use a TFC \cite{tfctdf} for each block with the same configuration used in \cite{tfctdf}, where 5 convolution layers with kernel size $3 \times 3$ are densely connected, and the growth rate \cite{densenet} is set to be 24. We set $\bar{C}$ to be 24 as in \cite{tfctdf}. We use 32 for the dimensionality of the embedding space of conditions (i.e., $\mathbb{R}^E$). We set $|\mathcal{I}_L|$, the number of latent instruments, to be 6.

In Table \ref{table:ablation}, we can observe a considerable performance degradation when employing the existing method (FiLM CUNet), compared to the \textit{dedicated} U-Net,  which is trained separately for each instrument with the same configuration.
Injecting TDF blocks to the baseline (FiLM CUNet + TDF) improves SDR by capturing the common frequency patterns. Replacing TDFs with LaSAFTs (FiLM CUNet + LaSAFT) significantly improves the average SDR score by 0.51dB, indicating that our LaSAFTs are more appropriate for multi-instrument tasks than TDFs.
Our proposed model (GPoCM CUNet + LaSAFT) outperforms the others, achieving comparable but slightly inferior results to dedicated models.

\begin{table}[]
\begin{tabular}{l|c|c|c|c|c}
\hline
model             & vocals & drums & bass & other & AVG \\ \hline
\textit{dedicated} \cite{tfctdf}     & 7.07   & 5.38  & 5.62 & 4.61  & 5.66    \\ \hline
FiLM CUNet     & 5.14   & 5.25  & 4.20 & 3.40  & 4.49    \\ 
 \quad + TDF    & 5.88   & 5.70  & 4.55 & 3.67  & 4.95    \\ 
 \quad + LaSAFT   & 6.74    & 5.64   & 5.13  & 4.32  & 5.46     \\ \hline
GPoCM CUNet    & 5.43 & 5.30 & 4.43 & 3.51  & 4.67     \\
 \quad + TDF  & 5.94   & 5.46  & 4.47 & 3.81  & 4.92    \\
 \quad + LaSAFT & \textbf{6.96}   & \textbf{5.84}  & \textbf{5.24} & \textbf{4.54}  & \textbf{5.64}    \\ \hline
\end{tabular}
    \caption{An ablation study: \textit{dedicated} means U-Nets for the single source separation, trained separately. FiLM CUNet refers the baseline in \S \ref{sec:baseline}. The last row is our proposed model. }
    \label{table:ablation}
\end{table}

Also, we compare against existing state-of-the-art models on the MUSDB18 benchmark. 
The first four rows of Table \ref{table:sota} show SDR scores of SOTA models. We take the SDR performance from the respective papers.
For fair comparison, we use a 9-blocked `GPoCM CUNet + LaSAFT' with the same frequency resolution as the other SOTA models \cite{dilatedlstm, d3net} (FFT window size = 4096). 
Our model yields comparable results against the existing methods and even outperforms the others on `vocals' and `other.'
LaSAFT's ability, which allows the proposed model to extract latent instrument-attentive frequency patterns, significantly improves the SDR of `other', since it contains various instruments such as piano and guitars.
Also, existing works \cite{tfctdf,phasen} have shown that FT-based methods are beneficial for voice separation, which explains our model's excellent SDR performance on vocals.


\begin{table}[]
\begin{tabular}{l|c|c|c|c|c}
\hline
model             & vocals & drums & bass & other & AVG \\ \hline
DGRU-DConv\cite{dilatedlstm} & 6.85 & 5.86 & 4.86 & 4.65 & 5.56 \\
Meta-TasNet\cite{meta}$^{*}$ & 6.40   & 5.91  & 5.58 & 4.19 & 5.52    \\ 
Nachmani\cite{Nachmani}$^{*}$ & 6.92 & 6.15 & \textbf{5.88} & 4.32 & 5.82 \\ 
D3Net \cite{d3net} & 7.24   & \textbf{7.01}  & 5.25 & 4.53 & \textbf{6.01}    \\ \hline
\textit{proposed} & \textbf{7.33}  & 5.68  & 5.63 & \textbf{4.87}  & 5.88    \\ \hline
\end{tabular}
    \caption{A comparison SDR performance of our models with other systems. `$*$' denotes model operating in time domain. }
    \label{table:sota}
\end{table}

\section{RELATION TO PRIOR WORK}
We first summarize the differences between the original CUNet \cite{cunet}, our baseline, and the proposed model as follows:
(1) our proposed model's U-Net is based on a generalized U-Net for source separation used in \cite{tfctdf}, but the U-Net of \cite{cunet} is more similar to the original U-Net \cite{medical_unet}, and (2) our baseline/proposed model applies FiLM/GPoCM to internal features in the decoder, but \cite{cunet} apply it in the encoder. The authors of \cite{cunet} tried to manipulate latent space in the encoder, assuming the decoder can perform as a general spectrogram generator, which is `shared' by different sources. 
However, we found that this approach is not practical since it makes the latent space (i.e., the decoder's input feature space) more discontinuous. Via preliminary experiments, we observed that applying FiLMs in the decoder was consistently better than applying FilMs in the encoder.

For multi-source separation, \cite{meta} employed meta-learning, which is similar to conditioning learning. It also has external networks that generate parameters for the target instrument. 
Whiling we focus on modulating internal representations with GPoCM, \cite{meta} focuses on generating parameters of the masking subnetwork.
Also, models proposed in \cite{meta,Nachmani} operate in time domain, while ours in time-frequency domain. 

While other multi-source separation methods \cite{dilatedlstm,d3net} estimate multiple sources simultaneously, we try to condition a shared U-Net. We expect that ours can be easily extended to more complicated tasks such as audio manipulation.

\label{sec:prior}

\section{Conclusion}
\label{sec:conclusion}

We propose LaSAFT that captures source-dependent frequency patterns by extending TDF to fit the multi-source task.
We also propose the GPoCM that modulates features more flexibly and expressively than FiLM.
We have shown that employing our LaSAFT and GPoCM in CUNet can significantly improve SDR performance.
In future work, we try to reduce the number of parameters and the memory usage of LaSAFT to consider more latent instruments.
Also, we extend the proposed model to audio manipulation tasks, where we can condition the model by providing various instructions.

\vfill\pagebreak

\bibliographystyle{IEEEbib}
\bibliography{strings,refs}

\end{document}